\documentclass[runningheads]{llncs}

\usepackage[numbers]{natbib} 

\usepackage{graphicx, subcaption} \graphicspath{ {Images/} }
\usepackage[pdftex, pdfpagelabels, hypertexnames, bookmarks, bookmarksnumbered, plainpages=false, naturalnames=false, pdfpagemode=UseOutlines]{hyperref}

\usepackage{amsmath}
\usepackage{booktabs}
\usepackage{microtype}
\usepackage{xcolor}
\usepackage[symbol]{footmisc}

\begin{document}

\title{Deep Active Lesion Segmentation}

\author{
Ali Hatamizadeh\inst{1*} \and
Assaf Hoogi\inst{2*} \and
Debleena Sengupta\inst{1} \and
Wuyue Lu\inst{1} \and \\
Brian Wilcox\inst{2} \and
Daniel Rubin\inst{2\dagger} \and
Demetri Terzopoulos\inst{1\dagger}
}

\footnotetext[1]{Co-primary authorship}
\footnotetext[2]{Co-senior authorship}

\authorrunning{A.~Hatamizadeh et al.}

\institute{Computer Science Department, University of California, Los Angeles, CA, USA \and
Department of Biomedical Data Science, Stanford University, Stanford, CA, USA}

\maketitle

\begin{abstract}
Lesion segmentation is an important problem in computer-assisted diagnosis that remains challenging due to the prevalence of low contrast, irregular boundaries that are unamenable to shape priors. We introduce Deep Active Lesion Segmentation (DALS), a fully automated segmentation framework that leverages the powerful nonlinear feature extraction abilities of fully Convolutional Neural Networks (CNNs) and the precise boundary delineation abilities of Active Contour Models (ACMs). Our DALS framework benefits from an improved level-set ACM formulation with a per-pixel-parameterized energy functional and a novel multiscale encoder-decoder CNN that learns an initialization probability map along with parameter maps for the ACM. We evaluate our lesion segmentation model on a new Multiorgan Lesion Segmentation (MLS) dataset that contains images of various organs, including brain, liver, and lung, across different imaging modalities---MR and CT. Our results demonstrate favorable performance compared to competing methods, especially for small training datasets. 

\textbf{Source code} : {\color{blue}{\href{https://github.com/ahatamiz/dals}{https://github.com/ahatamiz/dals}}}

\end{abstract}

\newcommand{\imgsize}{0.19}

\begin{figure*}[t]
\centering
\centerline{\makebox[\imgsize\textwidth]{(1) Brain MR} \makebox[\imgsize\textwidth]{(2) Liver MR} \makebox[\imgsize\textwidth]{(3) Liver CT} \makebox[\imgsize\textwidth]{(4) Lung CT}}
~\\[-5pt]
\subcaptionbox{Expert Manual}
{%
\includegraphics[width=\imgsize\textwidth,height=\imgsize\textwidth]{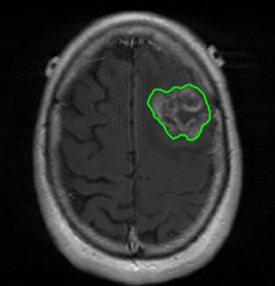} \hfill
\includegraphics[width=\imgsize\textwidth,height=\imgsize\textwidth]{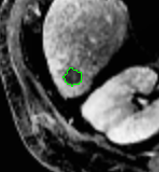} \hfill
\includegraphics[width=\imgsize\textwidth,height=\imgsize\textwidth,trim={270 80 230 100},clip]{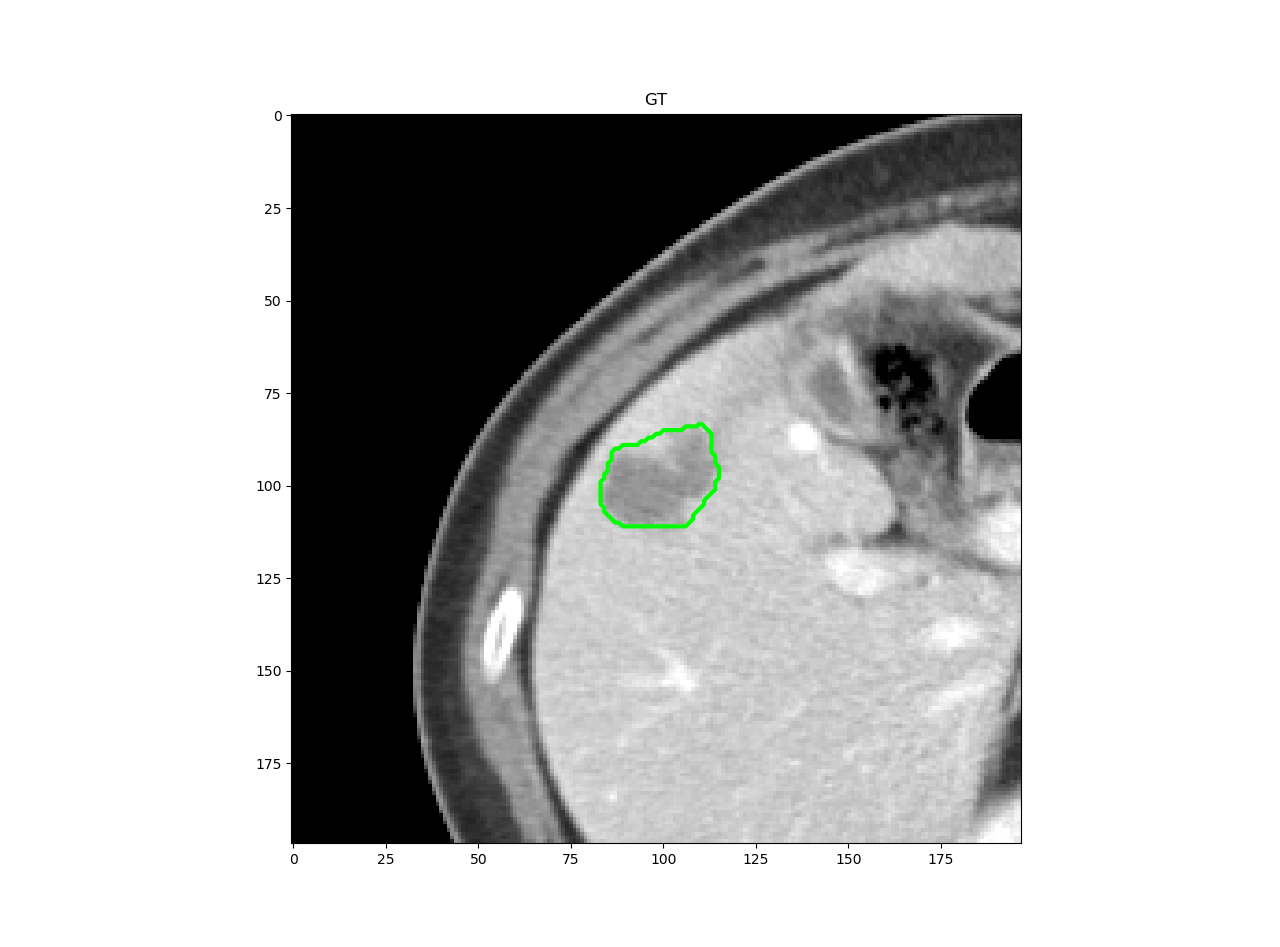} \hfill
\includegraphics[width=\imgsize\textwidth,height=\imgsize\textwidth,trim={270 80 230 100},clip]{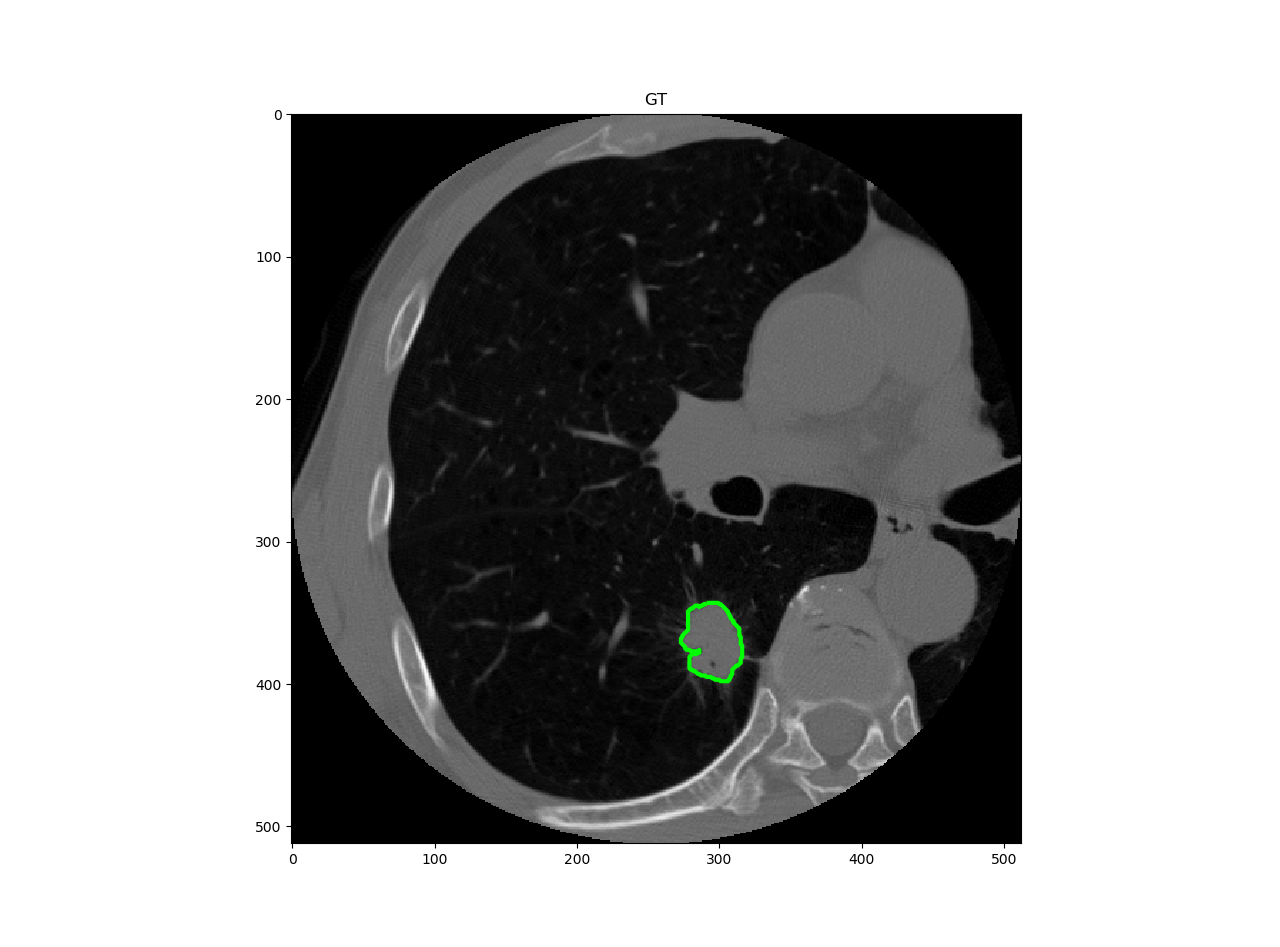}}\\[8pt]
\subcaptionbox{DALS Output}
{%
\includegraphics[width=\imgsize\textwidth,height=\imgsize\textwidth]{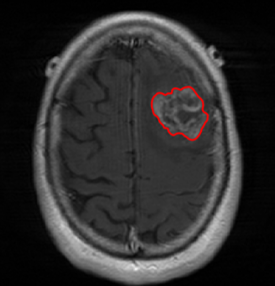} \hfill
\includegraphics[width=\imgsize\textwidth,height=\imgsize\textwidth]{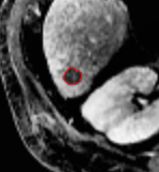} \hfill
\includegraphics[width=\imgsize\textwidth,height=\imgsize\textwidth,trim={270 80 230 100},clip]{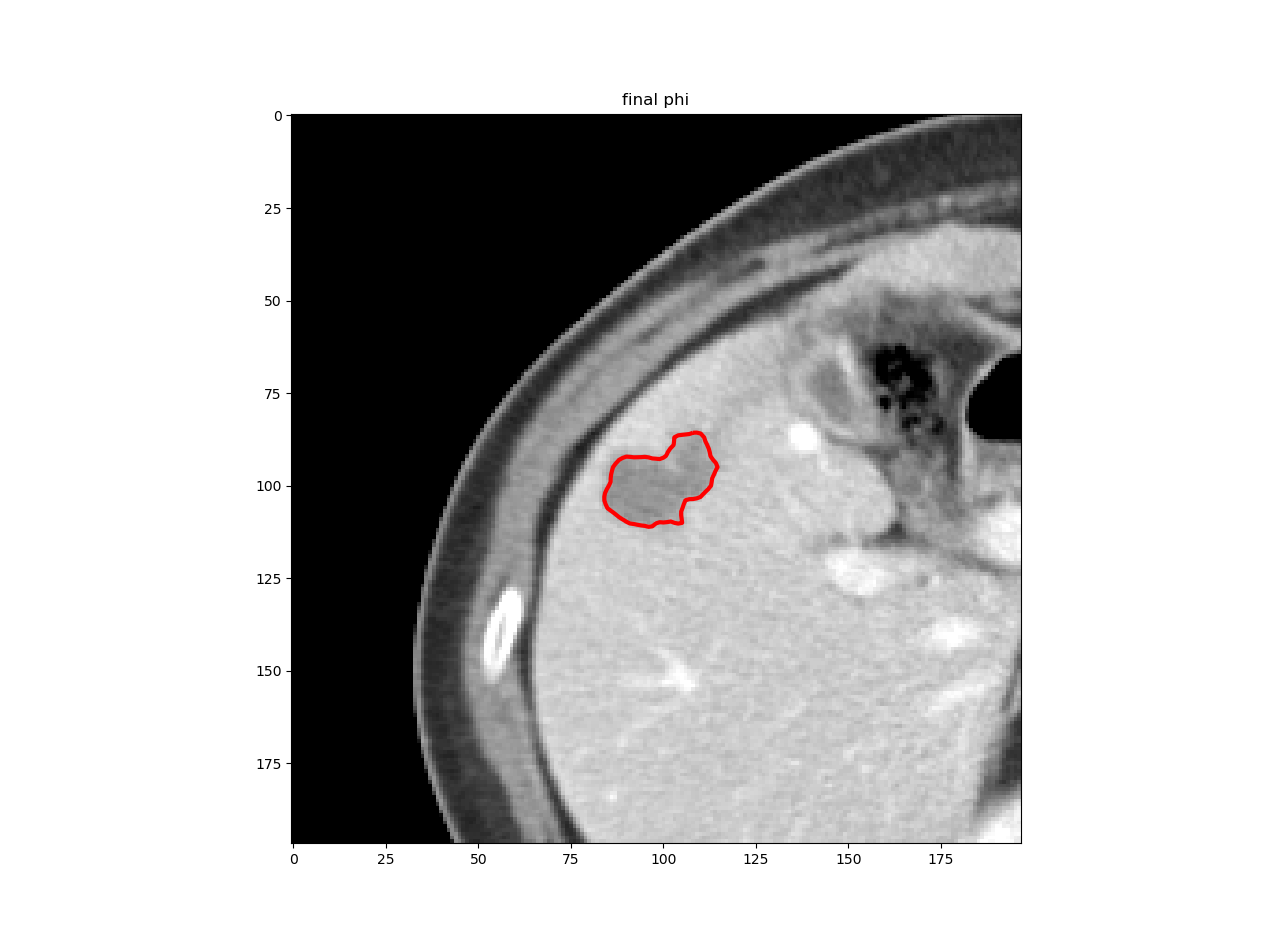} \hfill
\includegraphics[width=\imgsize\textwidth,height=\imgsize\textwidth,trim={270 80 230 100},clip]{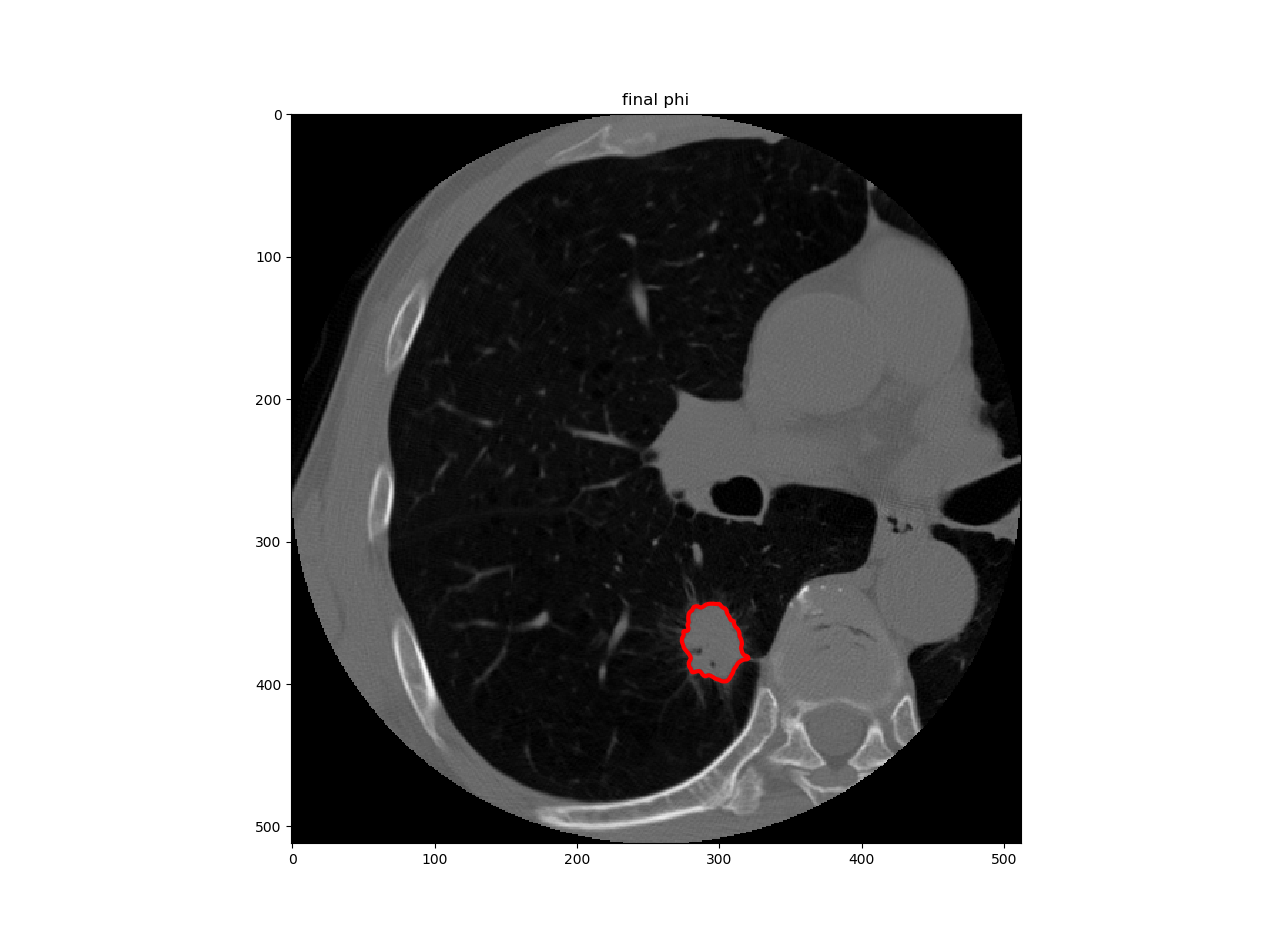}}\\[8pt]
\subcaptionbox{U-Net Output}
{%
\includegraphics[width=\imgsize\textwidth,height=\imgsize\textwidth]{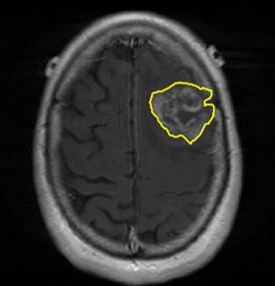} \hfill
\includegraphics[width=\imgsize\textwidth,height=\imgsize\textwidth]{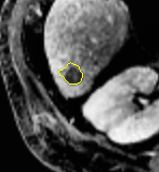} \hfill
\includegraphics[width=\imgsize\textwidth,height=\imgsize\textwidth,trim={270 80 230 100},clip]{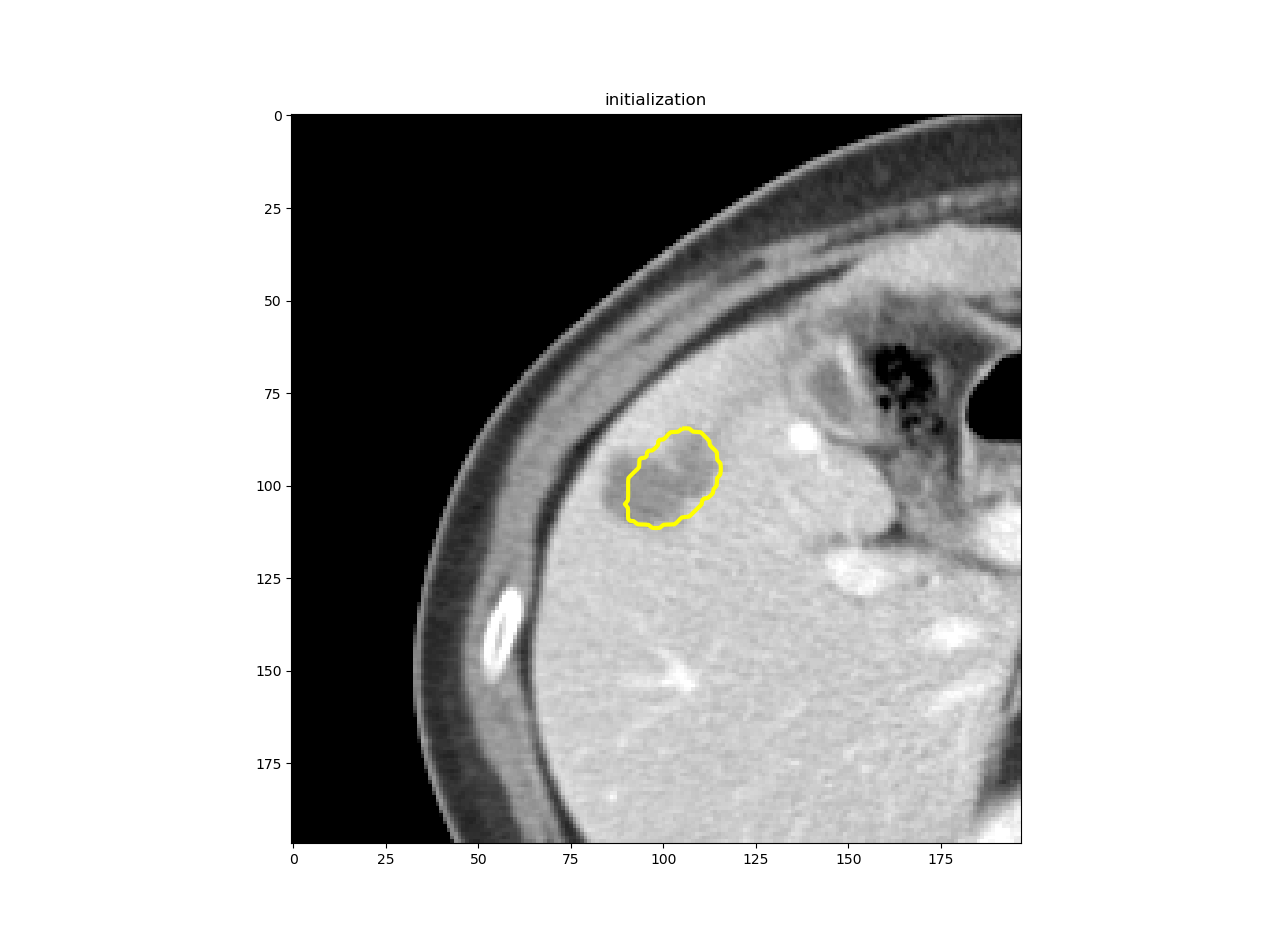} \hfill
\includegraphics[width=\imgsize\textwidth,height=\imgsize\textwidth,trim={270 80 230 100},clip]{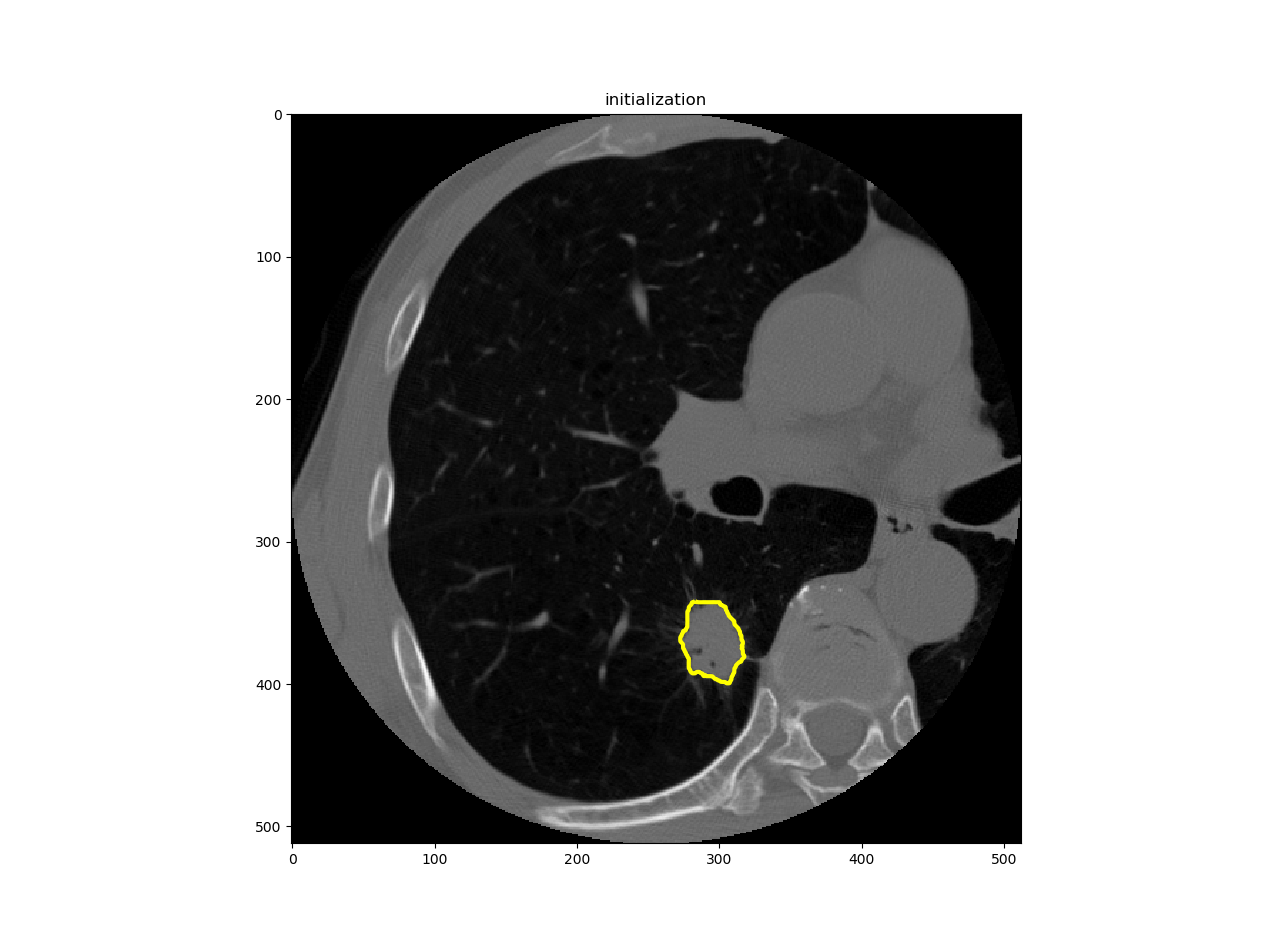}}
\caption{Segmentation comparison of (a) medical expert manual with (b) our DALS and (c) U-Net \citep{ronneberger2015u}, in (1) Brain MR, (2) Liver MR, (3) Liver CT, and (4) Lung CT images.}
\label{fig:comp1}
\end{figure*}

\section{Introduction}

Active Contour Models (ACMs) \citep{kass1988snakes} have been extensively applied to computer vision tasks such as image segmentation, especially for medical image analysis. ACMs leverage parametric (``snake'') or implicit (level-set) formulations in which the contour evolves by minimizing an associated energy functional, typically using a gradient descent procedure. In the level-set formulation,
this amounts to solving a partial differential equation (PDE) to evolve object boundaries that are able to handle large shape variations, topological changes, and intensity inhomogeneities.
Alternative approaches to image segmentation that are based on deep learning have recently been gaining in popularity. Fully Convolutional Neural Networks (CNNs) can perform well in segmenting images within datasets on which they have been trained \citep{ronneberger2015u,hatamizadeh2018automatic,hatamizadeh2019deep}, but they may lack robustness when cross-validated on other datasets. Moreover, in medical image segmentation, CNNs tend to be less precise in boundary delineation than ACMs.

In recent years, some researchers have sought to combine ACMs and deep learning approaches. \citet{hu2017deep} proposed a model in which the network learns a level-set function for salient objects; however, they predefined a fixed weighting parameter $\lambda$ with no expectation of optimality over all cases in the analyzed set of images.
\citet{marcos2018learning} combined CNNs and parametric ACMs for the segmentation of buildings in aerial images; however, their method requires manual contour initialization, fails to precisely delineate the boundary of complex shapes, and segments only single instances, all of which limit its applicability to lesion segmentation due to the irregular shapes of lesion boundaries and widespread cases of multiple lesions (e.g., liver lesions). 

We introduce a fully automatic framework for medical image segmentation that combines the strengths of CNNs and level-set ACMs to overcome their respective weaknesses. We apply our proposed Deep Active Lesion Segmentation (DALS) framework to the challenging problem of lesion segmentation in MR and CT medical images (Fig.~\ref{fig:comp1}), dealing with lesions of substantially different sizes within a single framework. In particular, our proposed encoder-decoder architecture learns to localize the lesion and generates an initial attention map along with associated parameter maps, thus instantiating a level-set ACM in which every location on the contour has local parameter values. We evaluate our lesion segmentation model on a new Multiorgan Lesion Segmentation (MLS) dataset that contains images of various organs, including brain, liver, and lung, across different imaging modalities---MR and CT. By automatically initializing and tuning the segmentation process of the level-set ACM, our DALS yields significantly more accurate boundaries in comparison to conventional CNNs and can reliably segment lesions of various sizes.

\section{Method}

\subsection{Level-Set Active Contour Model With Parameter Functions}

We introduce a generalization of the level-set ACM proposed by \citet{chan2001active}. Given an image $I(x,y)$, let $C(t) = \big\{(x, y) | \phi(x, y,t) = 0 \big\}$ be a closed time-varying contour represented in $\Omega\in R^2$ by the zero level set of the signed distance map $\phi(x,y,t)$. We select regions within a square window of size $s$ with a characteristic function $W_s$. The interior and exterior regions of $C$ are specified by the smoothed Heaviside function $H^I_\epsilon(\phi)$ and $H^E_\epsilon(\phi) = 1 - H^I_\epsilon(\phi)$, and the narrow band near $C$ is specified by the smoothed Dirac function $\delta_\epsilon(\phi)$. Assuming a uniform internal energy model \citep{chan2001active},  we follow Lankton \textit{et al.} \cite{lankton2008localizing} and define $m_1$ and $m_2$ as the mean intensities of $I(x,y)$ inside and outside $C$ and within $W_s$. Then, the energy functional associated with $C$ can be written as
\begin{equation}
\label{eq:f_pc}
\begin{split}
E(\phi)= \int_\Omega \delta_\epsilon(\phi(x,y,t)) \left(\mu|\nabla\phi(x,y,t)| + \int_\Omega W_s F(\phi(u,v,t)) \,du\,dv\right)\,dx\,dy,
\end{split}
\end{equation}
where $\mu$ penalizes the length of $C$ (we set $\mu=0.1$) and the energy density is
\begin{equation}
\label{eq:f_pc2}
\begin{split}
F(\phi) &= \lambda_1(u,v) (I(u,v)-m_1(x,y))^2 H^I_\epsilon(\phi)\\ &+ \lambda_2(u,v) (I(u,v)-m_2(x,y))^2 H^E_\epsilon(\phi).
\end{split}
\end{equation}

Note that to afford greater control over $C$, in (\ref{eq:f_pc2}) we have generalized the scalar parameter constants $\lambda_1$ and $\lambda_2$ used in \citep{chan2001active} to \emph{parameter functions} $\lambda_1(x,y)$ and $\lambda_2(x,y)$ over the image domain. Given an initial distance map $\phi(x,y,0)$ and parameter maps $\lambda_1(x,y)$ and $\lambda_2(x,y)$, the contour is evolved by numerically time-integrating, within a narrow band around $C$ for computational efficiency, the finite difference discretized Euler-Lagrange PDE for $\phi(x,y,t)$ (refer to \citep{chan2001active} and \citep{lankton2008localizing} for the details).

\subsection{CNN Backbone}

\begin{figure*}[t]
\includegraphics[width=\textwidth]{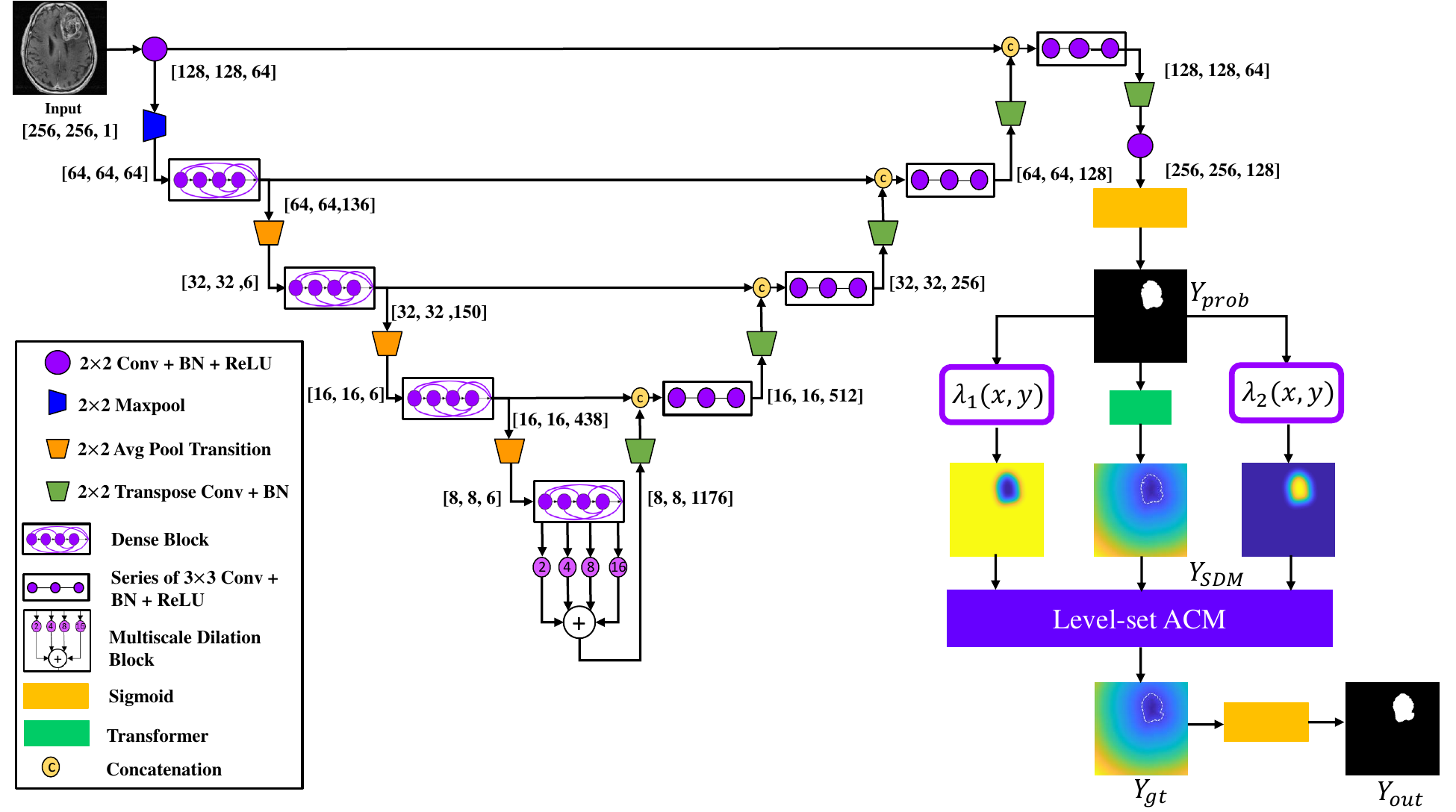}
\caption{The proposed DALS architecture. DALS is a fully automatic framework without the need for human supervision. The CNN initializes and guides the ACM by its learning local weighted parameters.}
\label{fig:dlac}
\end{figure*}

Our encoder-decoder is a fully convolutional architecture (Fig.~\ref{fig:dlac}) that is tailored and trained to estimate a probability map from which the initial distance function $\phi(x,y,0)$ of the level-set ACM and the functions $\lambda_1(x,y)$ and $\lambda_2(x,y)$ are computed. In each dense block of the encoder, a composite function of batch normalization, convolution, and ReLU is applied to the concatenation of all the feature maps $[x_0, x_1, \dots , x_{l-1}]$ from layers 0 to $l-1$ with the feature maps produced by the current block.  This concatenated result is passed through a transition layer before being fed to successive dense blocks. The last dense block in the encoder is fed into a custom multiscale dilation block with 4 parallel convolutional layers with dilation rates of 2, 4, 8, and 16. Before being passed to the decoder, the output of the dilated convolutions are then concatenated to create a multiscale representation of the input image thanks to the enlarged receptive field of its dilated convolutions. This, along with dense connectivity, assists in capturing local and global context for highly accurate lesion localization.

\subsection{The DALS Framework}

Our DALS framework is illustrated in Fig.~\ref{fig:dlac}. 
The boundaries of the segmentation map generated by the encoder-decoder are fine-tuned by the level-set ACM that takes advantage of information in the CNN maps to set the per-pixel parameters and initialize the contour.
The input image is fed into the encoder-decoder, which localizes the lesion and, after $1\times1$ convolutional and sigmoid layers, produces the initial segmentation probability map $Y_\textit{prob}(x,y)$, which specifies the probability that any point $(x,y)$ lies in the interior of the lesion. The Transformer converts $Y_\textit{prob}$ to a Signed Distance Map (SDM) $\phi(x,y,0)$ that initializes the level-set ACM. Map $Y_\textit{prob}$ is also utilized to estimate the parameter functions $\lambda_1(x,y)$ and $\lambda_2(x,y)$ in the energy functional (\ref{eq:f_pc}). Extending the approach of \citet{hoogi2017adaptive}, the $\lambda$ functions in Fig.~\ref{fig:dlac} are chosen as follows:
\begin{equation}
\label{eq:lambdas}
\lambda_1(x,y)=\exp\left(\frac{2-Y_\textit{prob}(x,y)}{1+Y_\textit{prob}(x,y)}\right); \quad
\lambda_2(x,y)=\exp\left(\frac{1+Y_\textit{prob}(x,y)}{2-Y_\textit{prob}(x,y)}\right).
\end{equation}
The exponential amplifies the range of values that the functions can take. These computations are performed for each point on the zero level-set contour $C$. During training, $Y_\textit{prob}$ and the ground truth map $Y_\textit{gt}(x,y)$ are fed into a Dice loss function and the error is back-propagated accordingly. During inference, a forward pass through the encoder-decoder and level-set ACM results in a final SDM, which is converted back into a probability map by a sigmoid layer, thus producing the final segmentation map $Y_\textit{out}(x,y)$.

\paragraph{Implementation Details:}

DALS is implemented in Tensorflow. We trained it on an NVIDIA Titan XP GPU and an Intel® Core™ i7-7700K CPU @ 4.20GHz. All the input images were first normalized and resized to a predefined size of $256\times256$ pixels. The size of the mini-batches is set to 4, and the Adam optimization algorithm was used with an initial learning rate of 0.001 that decays by a factor of 10 every 10 epochs. The entire inference time for DALS takes $1.5$ seconds. All model performances were evaluated by using the Dice coefficient, Hausdorff distance, and BoundF. 

\section{Multiorgan Lesion Segmentation (MLS) Dataset}

As shown in Table~\ref{table:stanford-dataset}, the MLS dataset includes images of highly diverse lesions in terms of size and spatial characteristics such as contrast and homogeneity. The liver component of the dataset consists of 112 contrast-enhanced CT images of liver lesions (43 hemangiomas, 45 cysts, and 24 metastases) with a mean lesion radius of 20.483 $\pm$ 10.37 pixels and 164 liver lesions from 3T gadoxetic acid enhanced MRI scans (one or more LI-RADS (LR), LR-3, or LR-4 lesions) with a mean lesion radius of 5.459 $\pm$ 2.027 pixels. The brain component consists of 369 preoperative and pretherapy perfusion MR images with a mean lesion radius of 17.42 $\pm$ 9.516 pixels. The lung component consists of 87 CT images with a mean lesion radius of 15.15 $\pm$ 5.777 pixels. For each component of the MLS dataset, we used 85\% of its images for training, 10\% for testing, and 5\% for validation.  

\begin{table*}[t] \setlength{\tabcolsep}{4pt}
\caption{MLS dataset statistics. GC:~Global Contrast; GH:~Global Heterogeneity.}
\label{table:stanford-dataset}
\resizebox{\linewidth}{!}{
\begin{tabular}{llcccccccr}
\toprule
Organ        & Modality    & \# Samples  & $\textrm{Mean}_\textrm{GC}$ & $\textrm{Var}_\textrm{GC}$ & $\textrm{Mean}_\textrm{GH}$ & $\textrm{Var}_\textrm{GH}$ & Lesion Radius (pixels)\\
\midrule
Brain &    MRI      &    369     &    0.56    & 0.029  & 0.907  & 0.003 &  17.42 $\pm$ 9.516    \\
Lung  &    CT      &    87      &    0.315   & 0.002  & 0.901  & 0.004 &  15.15 $\pm$ 5.777    \\
Liver &    CT      &    112     &    0.825   & 0.072  & 0.838  & 0.002 &  20.483 $\pm$ 10.37    \\
Liver &    MRI      &    164     &    0.448   & 0.041  & 0.891  & 0.003 &   5.459  $\pm$ 2.027  \\
\bottomrule
\end{tabular}
}
\end{table*}

\section{Results and Discussion}

\paragraph{Algorithm Comparison:}

\begin{figure*}[t]
    \centering
    \includegraphics[width=0.49\textwidth,trim={36 6 0 0},clip]{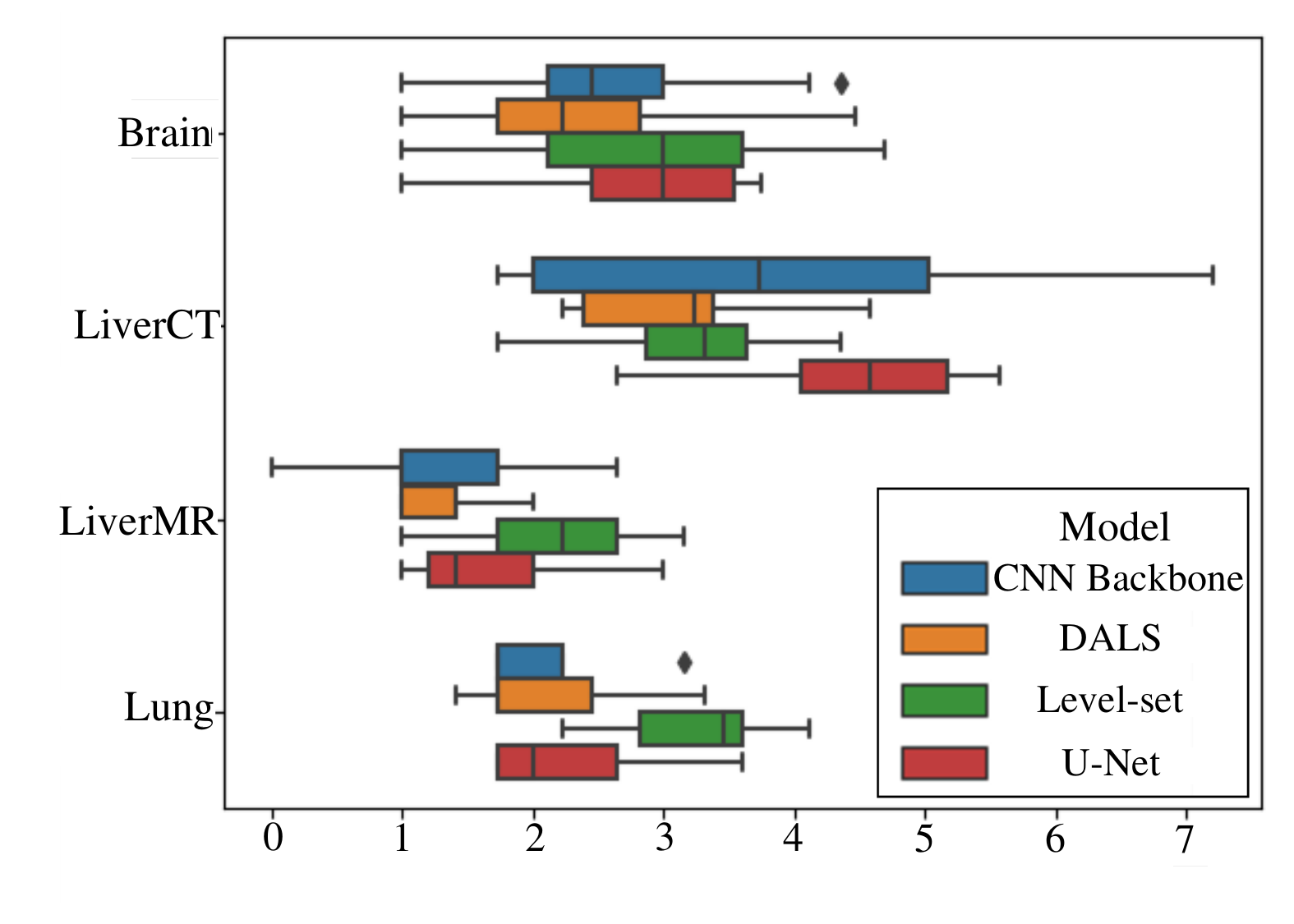}
    \hfill
    \includegraphics[width=0.48\textwidth,trim={36 -5 0 0},clip]{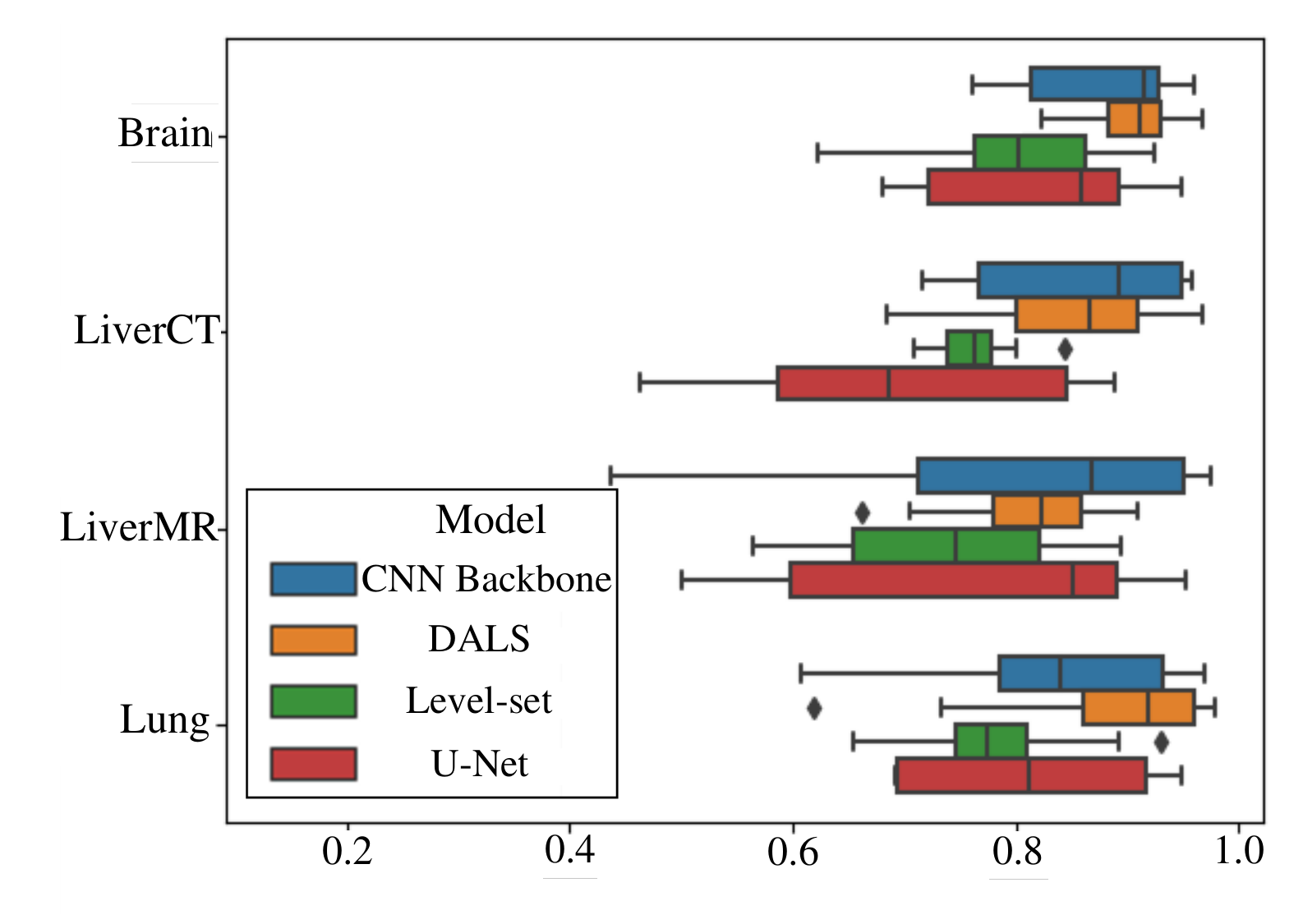}
    \makebox[0.52\linewidth]{(a)} \hfill \makebox[0.40\linewidth]{(b)} \hfill
    \caption{Box and whisker plots of: (a) Dice score; (b) Hausdorff distance.}
    \label{fig:box_plots}
\end{figure*}

We have quantitatively compared our DALS against U-Net \citep{ronneberger2015u} and manually-initialized level-set ACM with scalar $\lambda$ parameter constants as well as its backbone CNN. The evaluation metrics for each organ are reported in Table~\ref{table:datasets-perf} and box and whisker plots are shown in Fig.~\ref{fig:box_plots}. Our DALS achieves superior accuracies under all metrics and in all datasets. Furthermore, we evaluated the statistical significance of our method by applying a Wilcoxon paired test on the calculated Dice results. Our DALS performed significantly better than the U-Net ($p<0.001$), the manually-initialized ACM ($p<0.001$), and DALS's backbone CNN on its own ($p<0.005$).

\begin{table*}[t] \setlength{\tabcolsep}{4pt}
\caption{Segmentation metrics for model evaluations. Box and whisker plots are shown in Fig.~\ref{fig:box_plots}. CI denotes the confidence interval.}
\label{table:datasets-perf}

\hspace{-3mm}\resizebox{1.05\linewidth}{!}{
\begin{tabular}{lccccc|ccccc}
\toprule
Dataset:    & \multicolumn{5}{c}{Brain MR} & \multicolumn{5}{|c}{Lung CT}\\
\midrule
Model    &   Dice   &   CI &  Hausdorff &   CI   &   BoundF &
Dice   &    CI &  Hausdorff   &   CI   &   BoundF\\
\midrule

U-Net   & 0.776 $\pm$ 0.214 &	0.090 &	2.988 $\pm$  1.238 &	0.521 &	0.826  & 0.817 $\pm$ 0.098 &	0.0803 &	2.289 $\pm$ 0.650 &	0.53  &	0.898 \\

CNN Backbone  &   0.824 $\pm$ 0.193 &	0.078 &	2.755 $\pm$ 1.216 &	0.49 &	0.891 & 0.822 $\pm$ 0.115 &	0.0944 &	2.254 $\pm$  0.762&	0.6218 &	0.900 \\ 

Level-set  &   0.796 $\pm$ 0.095	& 0.038	&2.927 $\pm$ 0.992&	0.400 &	0.841 & 0.789 $\pm$ 0.078	& 0.064	& 3.27 $\pm$ 0.553 &	0.4514 &	0.879	 \\[2pt]

DALS     & \textbf{0.888 $\pm$  0.0755}  &	\textbf{0.03} &	\textbf{2.322 $\pm$ 0.824} &	\textbf{0.332}  &	\textbf{0.944} &	 \textbf{0.869 $\pm$ 0.113} & \textbf{0.092} &	\textbf{2.095 $\pm$ 0.623} &	\textbf{0.508}  &	\textbf{0.937} \\
\bottomrule
\end{tabular}} \\[3pt]

\hspace{-3mm}\resizebox{1.05\linewidth}{!}{
\begin{tabular}{lccccc|ccccc}
\toprule
Dataset:    & \multicolumn{5}{c}{Liver MR} & \multicolumn{5}{|c}{Liver CT}\\
\midrule
Model    &   Dice   &   CI &  Hausdorff   &   CI  &   BoundF &
Dice  &    CI &  Hausdorff  &   CI   &   BoundF\\
\midrule

U-Net   & 0.769 $\pm$ 0.162 &	0.093 &	1.645 $\pm$  0.598 &	0.343  &	0.92 &  0.698 $\pm$  0.149 &	0.133 &	4.422 $\pm$ 0.969  &	0.866 &	0.662\\

CNN Backbone  &   0.805 $\pm$ 0.193 &	0.11 &	1.347 $\pm$ 0.671 &	0.385 &	0.939 & 0.801 $\pm$ 0.178 &	0.159 &	3.813 $\pm$  1.791 &	1.6  &	0.697 \\

Level-set  &   0.739 $\pm$ 0.102 &	0.056 &	2.227 $\pm$ 0.576 & 	0.317  &	0.954 &	0.765 $\pm$ 0.039 &	0.034 &	3.153 $\pm$ 0.825 &	0.737  &	0.761 \\ [2pt]

DALS     & \textbf{0.894 $\pm$  0.0654} &	\textbf{0.036} &	\textbf{1.298 $\pm$ 0.434}	& \textbf{0.239}  &	\textbf{0.987} & \textbf{0.846 $\pm$  0.090} &	\textbf{0.0806} &	\textbf{3.113 $\pm$ 0.747} &	\textbf{0.667}  &	\textbf{0.773} \\
\bottomrule
\end{tabular}}
\bigskip 
\end{table*}

\begin{figure*}[t]
\centering
\subcaptionbox{Brain MR}
{%
\includegraphics[width=0.134\textwidth,height=0.134\textwidth,trim={000 000 000 000},clip]{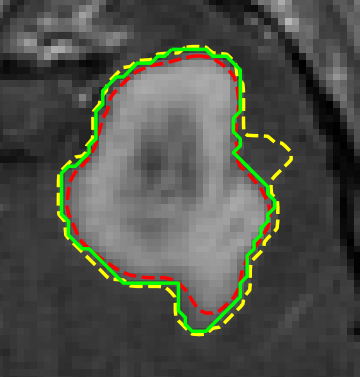} \hfill
\includegraphics[width=0.134\textwidth,height=0.134\textwidth,trim={150 120 150 120},clip]{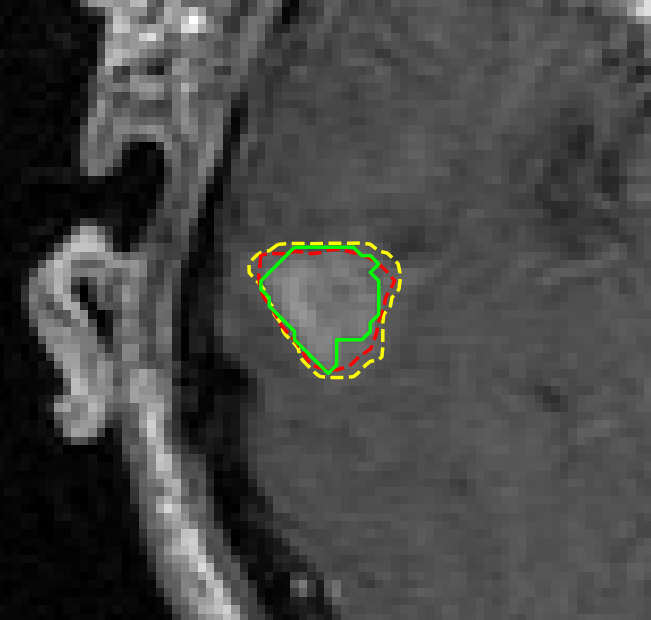} \hfill
\includegraphics[width=0.134\textwidth,height=0.134\textwidth,trim={120 170 180 170},clip]{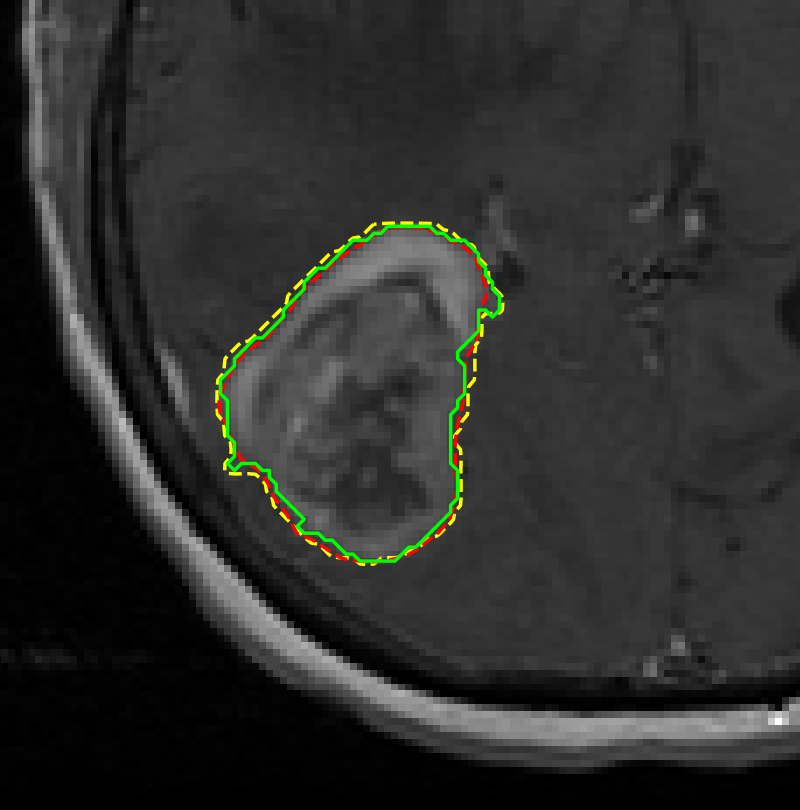} \hfill
\includegraphics[width=0.134\textwidth,height=0.134\textwidth,trim={150 120 150 120},clip]{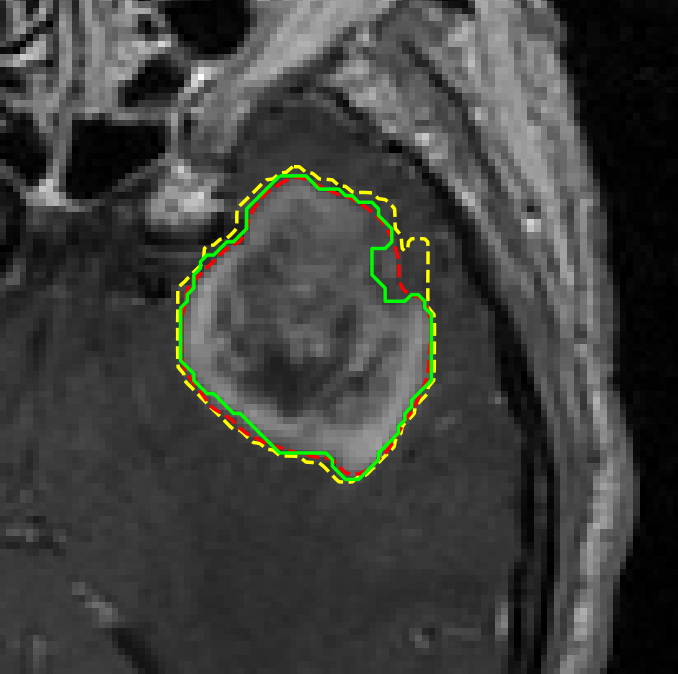} \hfill
\includegraphics[width=0.134\textwidth,height=0.134\textwidth,trim={150 120 150 120},clip]{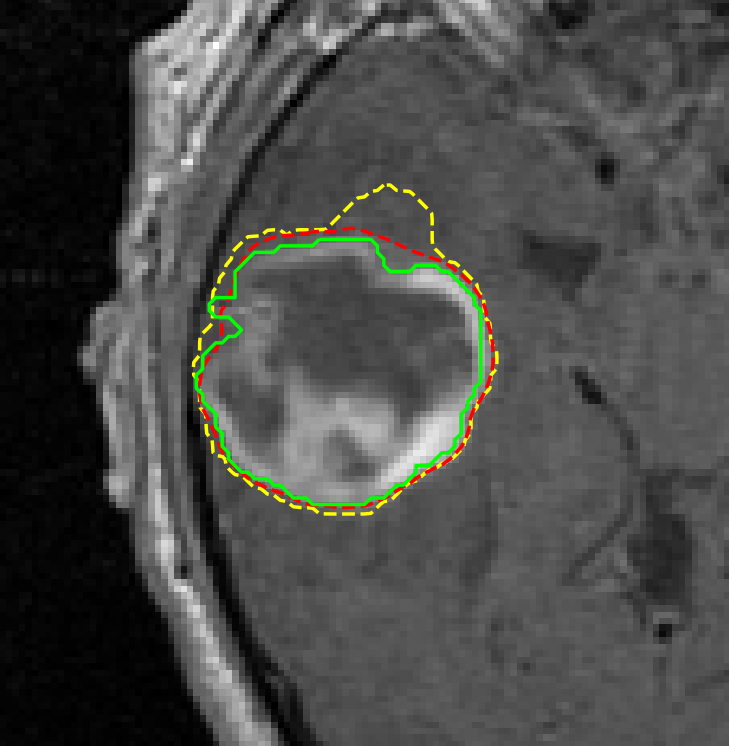} \hfill
\includegraphics[width=0.134\textwidth,height=0.134\textwidth,trim={150 120 150 120},clip]{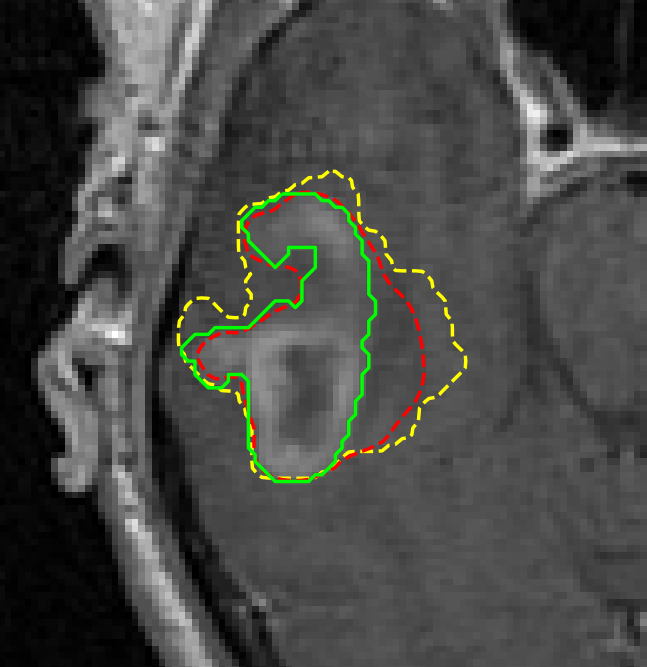} \hfill
\includegraphics[width=0.134\textwidth,height=0.134\textwidth,trim={110 120 120 120},clip]{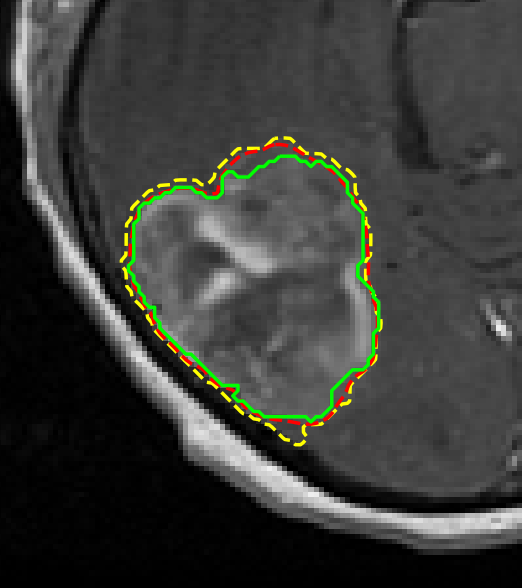}}\\[3pt]
\subcaptionbox{Liver CT}
{%
\includegraphics[width=0.134\textwidth,height=0.134\textwidth,trim={150 120 150 120},clip]{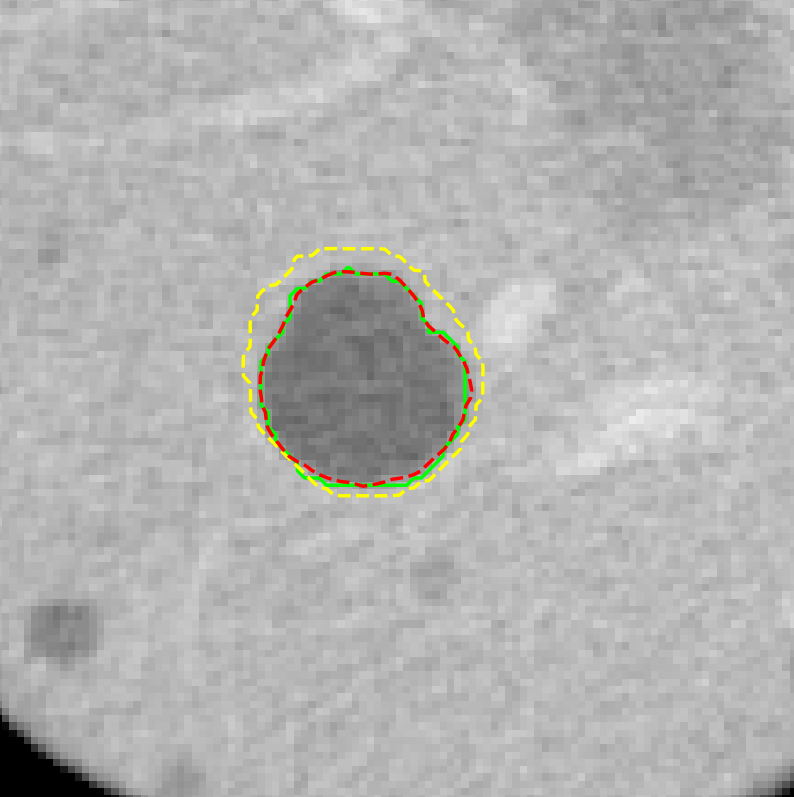} \hfill
\includegraphics[width=0.134\textwidth,height=0.134\textwidth,trim={150 120 150 120},clip]{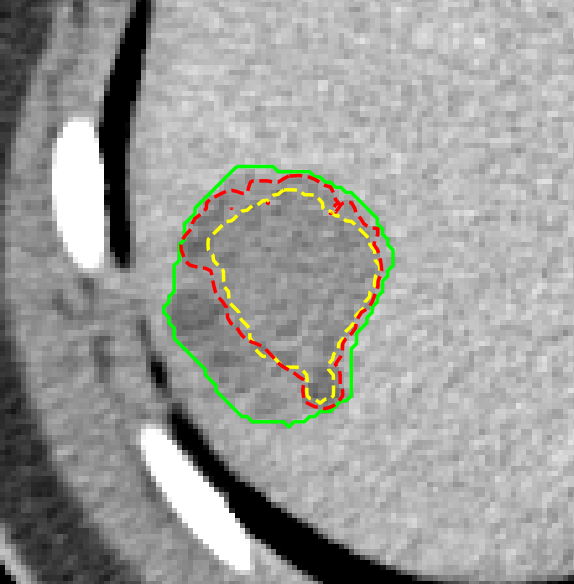} \hfill
\includegraphics[width=0.134\textwidth,height=0.134\textwidth,trim={150 120 150 120},clip]{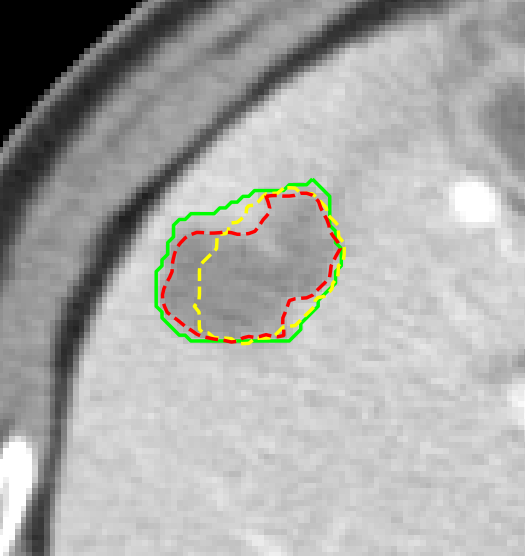} \hfill
\includegraphics[width=0.134\textwidth,height=0.134\textwidth,trim={150 120 150 120},clip]{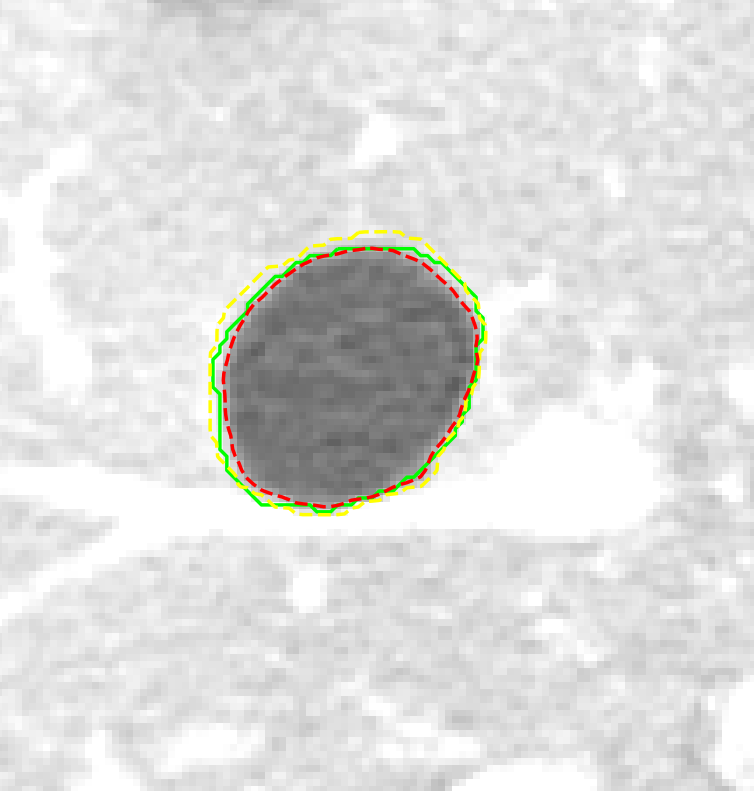} \hfill
\includegraphics[width=0.134\textwidth,height=0.134\textwidth,trim={150 120 150 120},clip]{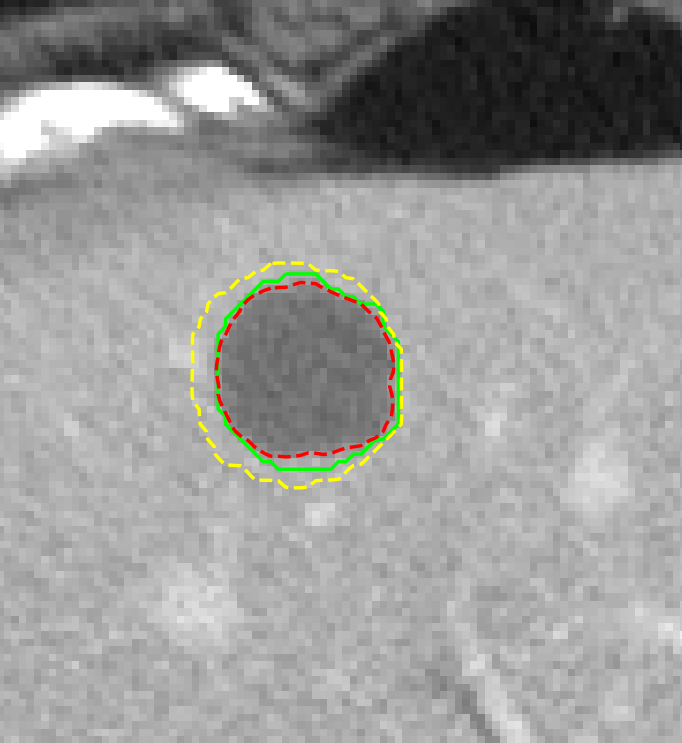} \hfill
\includegraphics[width=0.134\textwidth,height=0.134\textwidth,trim={110 120 130 120},clip]{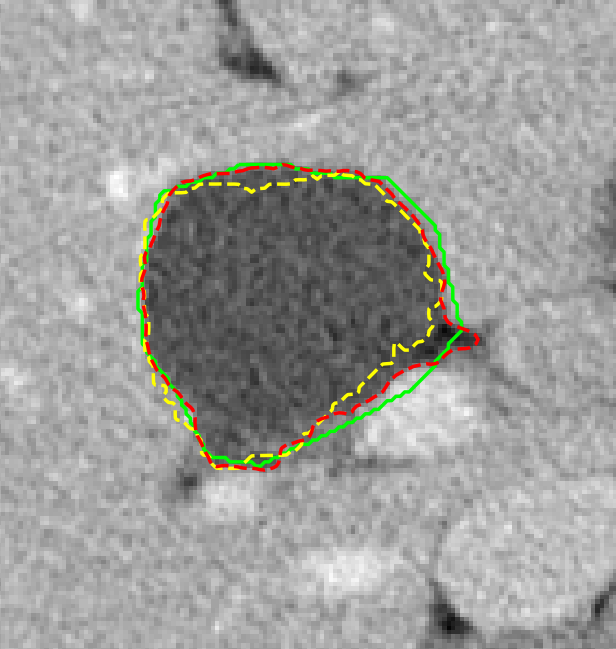} \hfill
\includegraphics[width=0.134\textwidth,height=0.134\textwidth,trim={150 120 150 120},clip]{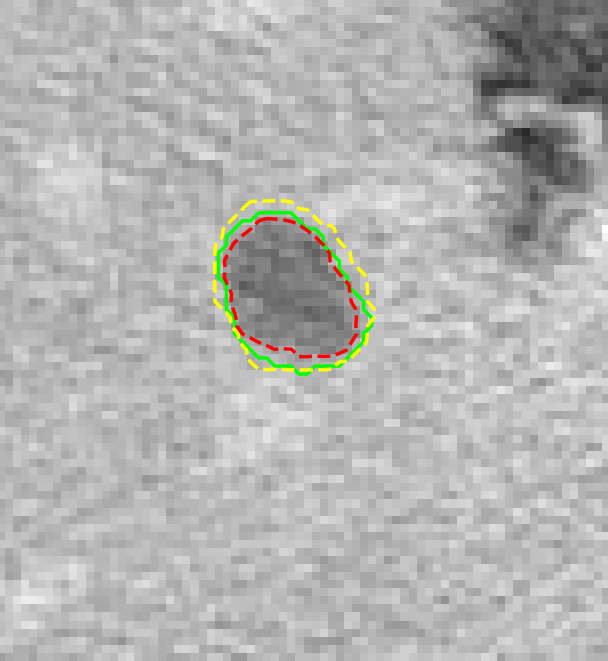}}\\[3pt]
\subcaptionbox{Liver MR}
{%
\includegraphics[width=0.134\textwidth,height=0.134\textwidth,trim={150 120 150 120},clip]{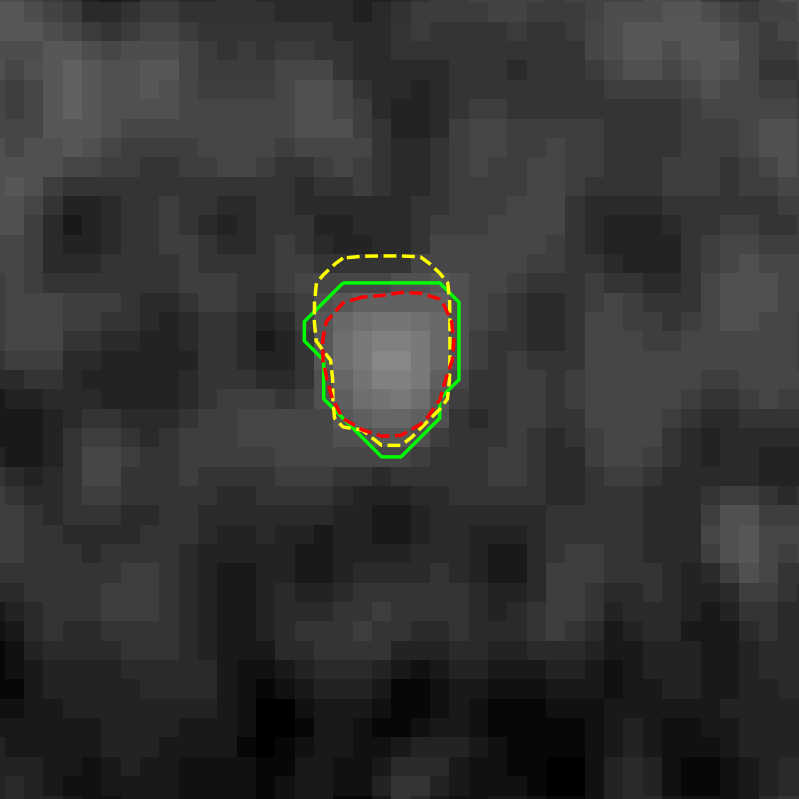} \hfill
\includegraphics[width=0.134\textwidth,height=0.134\textwidth,trim={150 120 150 120},clip]{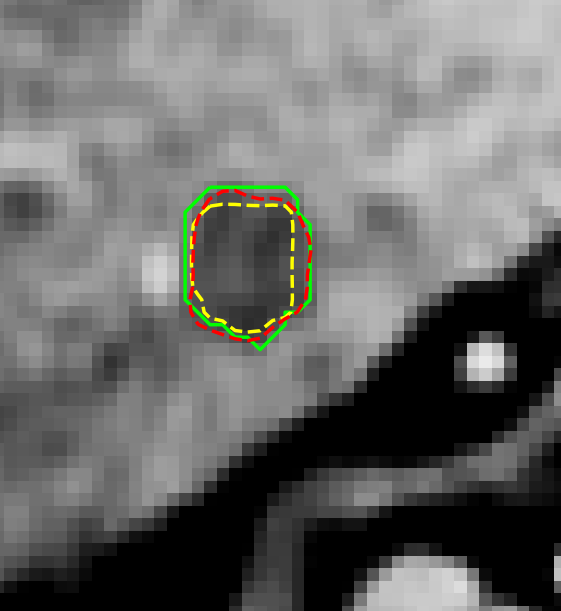} \hfill
\includegraphics[width=0.134\textwidth,height=0.134\textwidth,trim={150 120 150 120},clip]{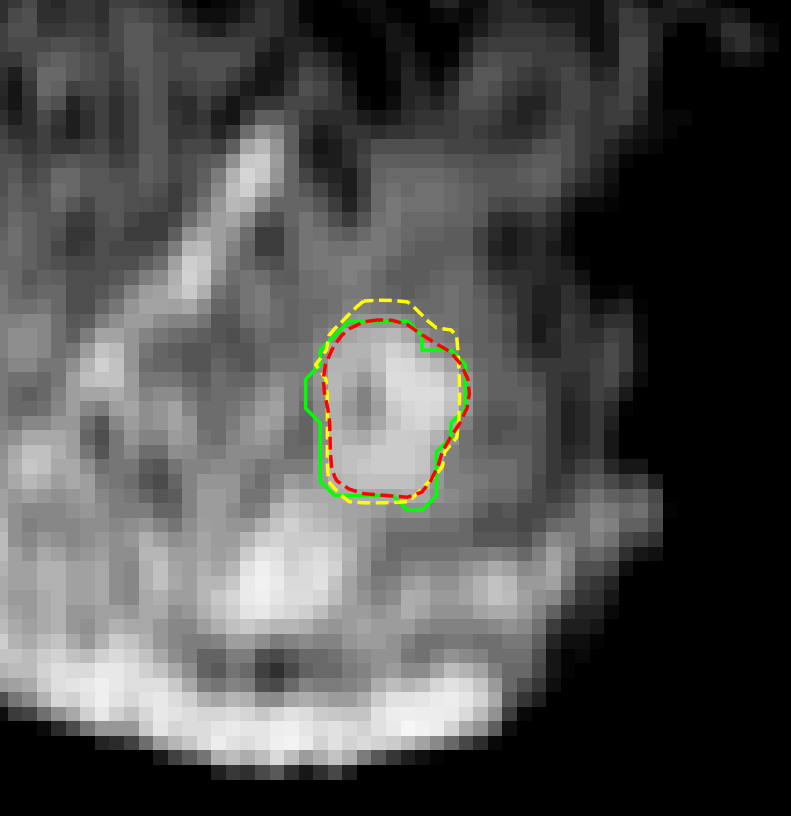} \hfill
\includegraphics[width=0.134\textwidth,height=0.134\textwidth,trim={150 120 150 120},clip]{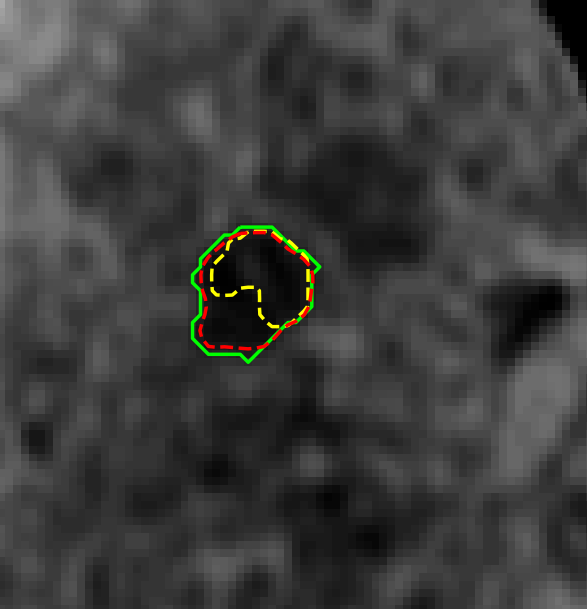} \hfill
\includegraphics[width=0.134\textwidth,height=0.134\textwidth,trim={150 120 150 120},clip]{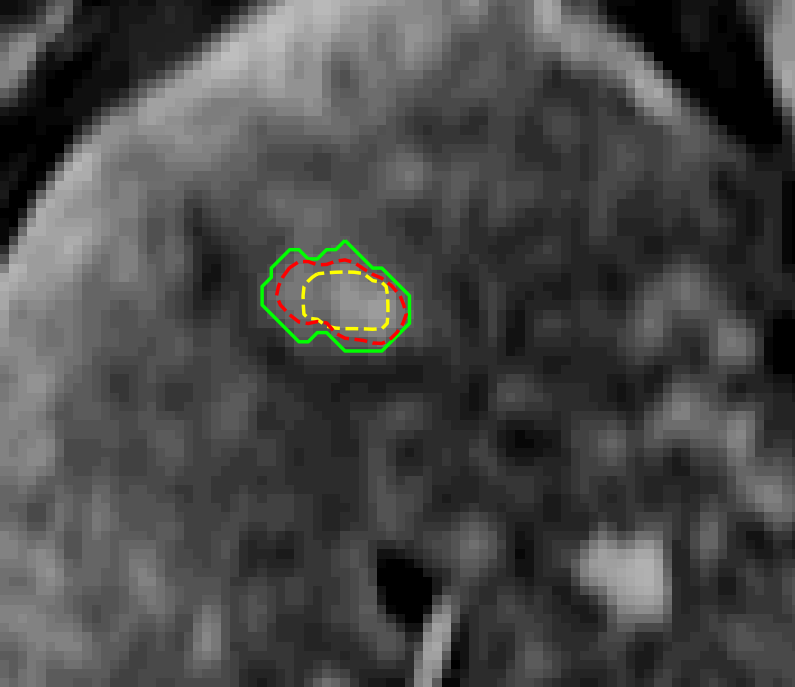} \hfill
\includegraphics[width=0.134\textwidth,height=0.134\textwidth,trim={150 120 150 120},clip]{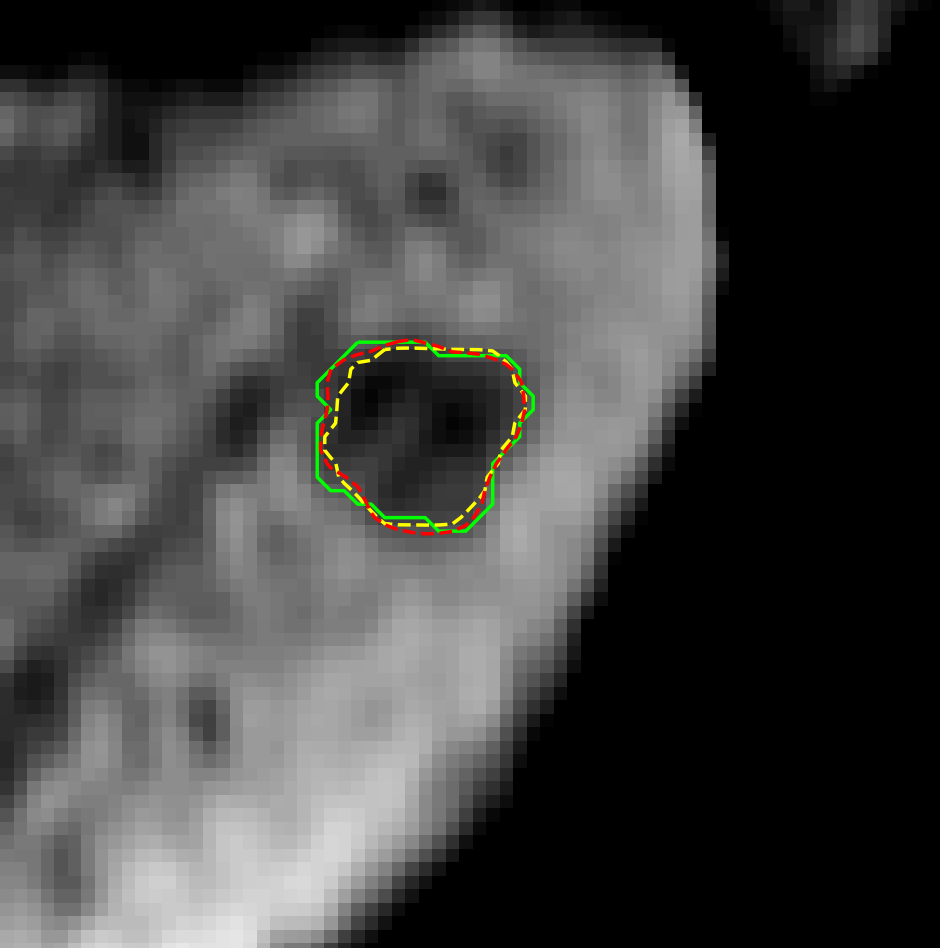} \hfill
\includegraphics[width=0.134\textwidth,height=0.134\textwidth,trim={150 120 150 120},clip]{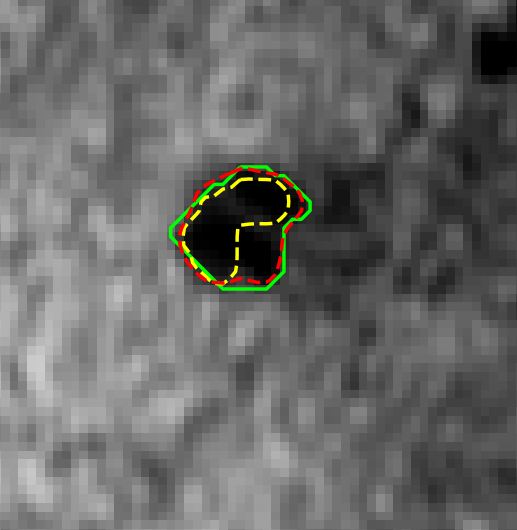}}\\[3pt]
\subcaptionbox{Lung CT}
{%
\includegraphics[width=0.134\textwidth,height=0.134\textwidth,trim={150 120 150 120},clip]{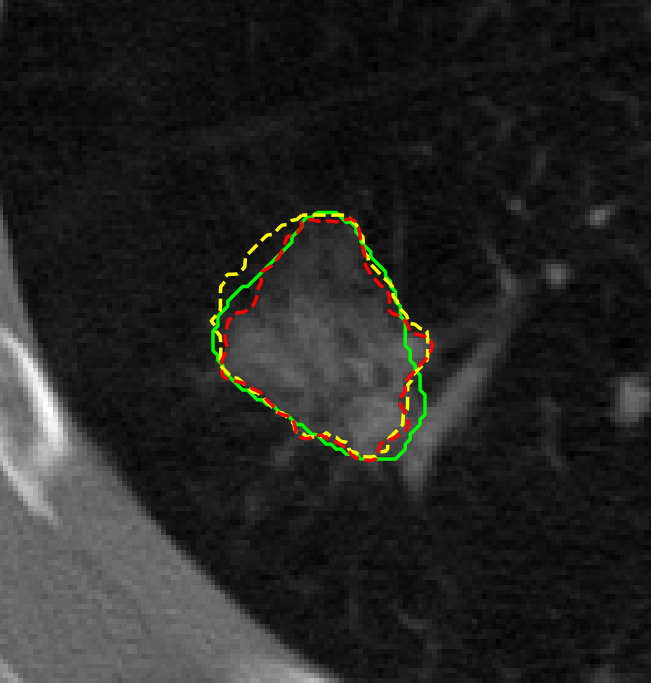} \hfill
\includegraphics[width=0.134\textwidth,height=0.134\textwidth,trim={150 120 150 120},clip]{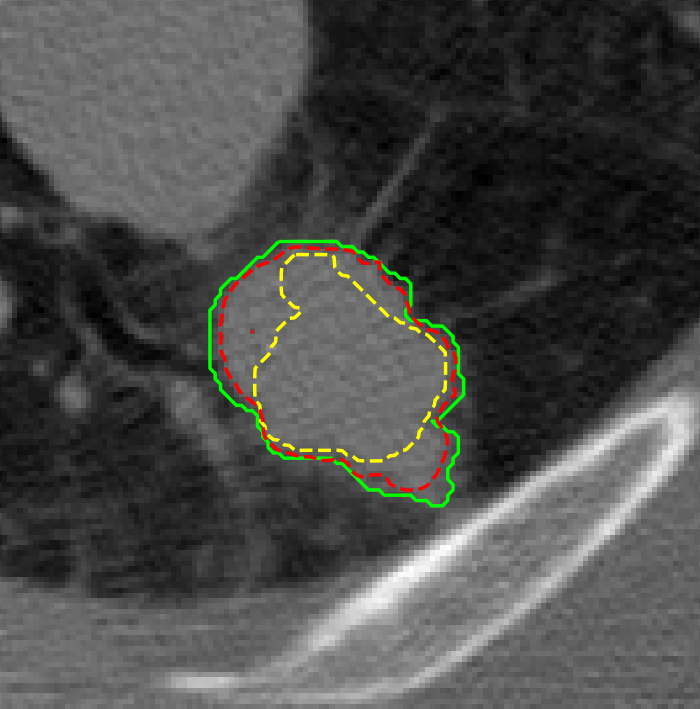} \hfill
\includegraphics[width=0.134\textwidth,height=0.134\textwidth,trim={150 120 150 120},clip]{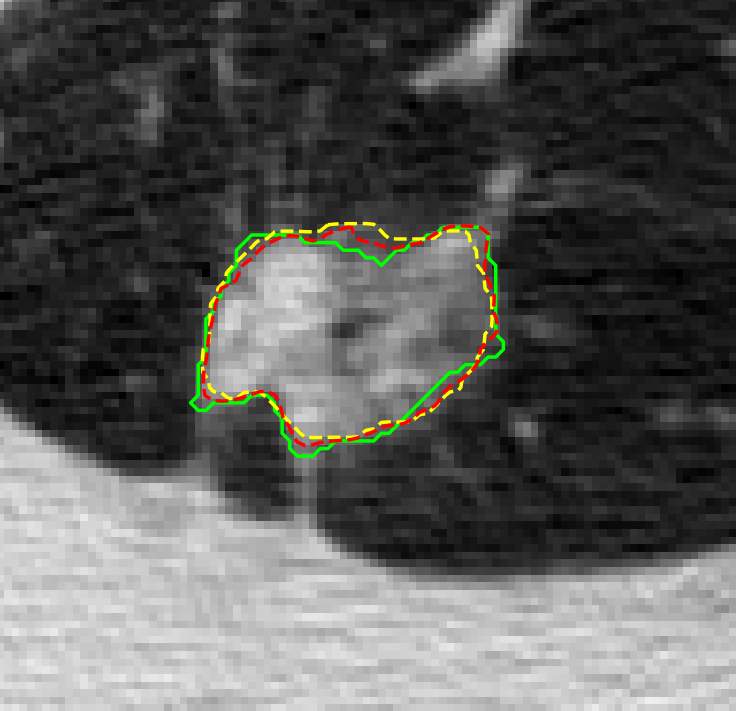} \hfill
\includegraphics[width=0.134\textwidth,height=0.134\textwidth,trim={150 120 150 120},clip]{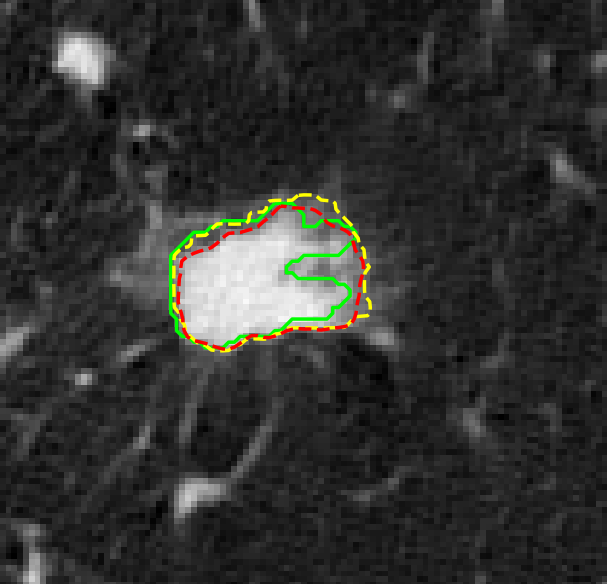} \hfill
\includegraphics[width=0.134\textwidth,height=0.134\textwidth,trim={150 120 150 120},clip]{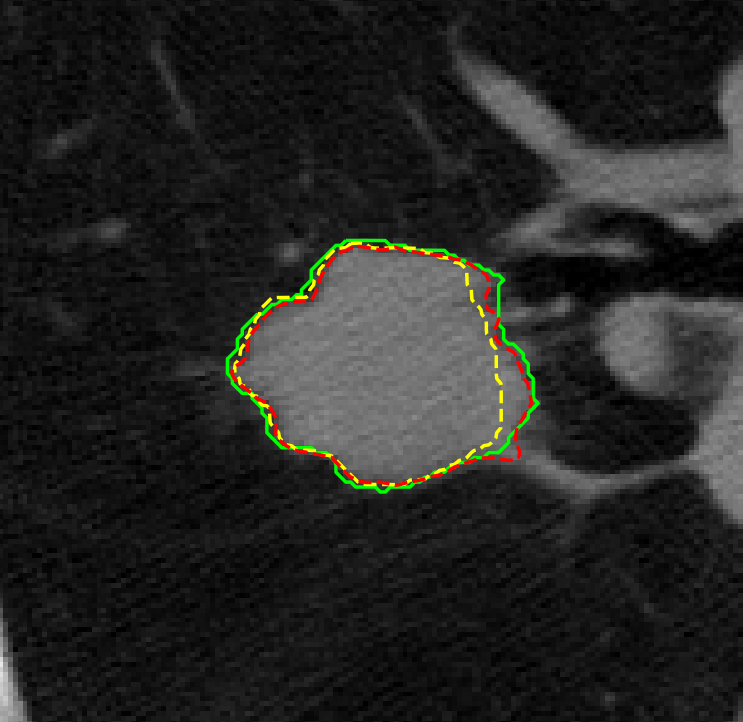} \hfill
\includegraphics[width=0.134\textwidth,height=0.134\textwidth,trim={150 120 150 120},clip]{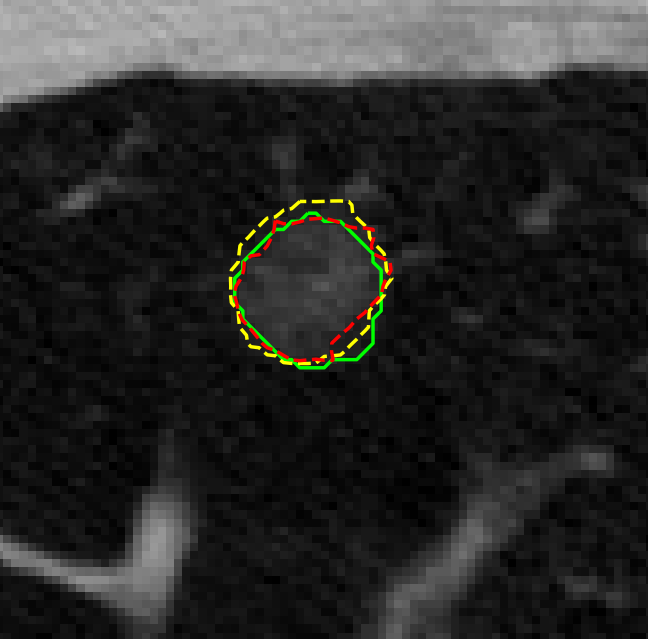} \hfill
\includegraphics[width=0.134\textwidth,height=0.134\textwidth,trim={150 120 150 120},clip]{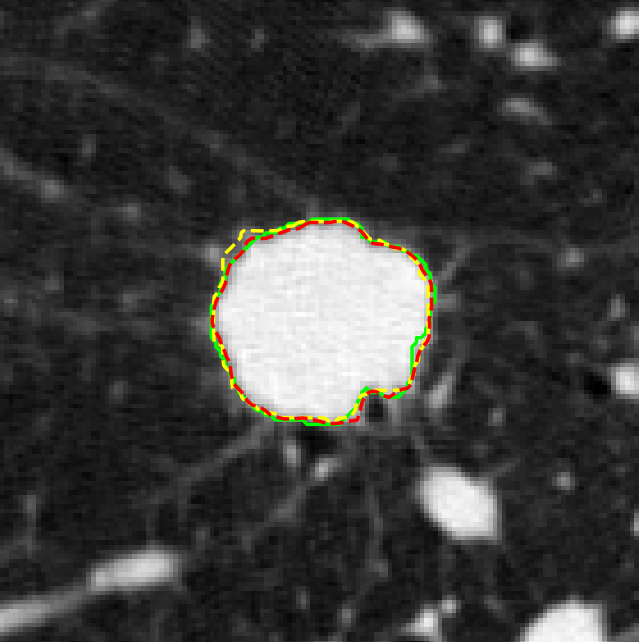}}
\caption{Comparison of the output segmentation of our DALS (red) against the U-Net \citep{ronneberger2015u} (yellow) and manual ``ground truth'' (green) segmentations on images of Brain MR, Liver CT, Liver MR, and Lung CT on the MLS test
set.}
\label{fig:result-comp}
\end{figure*}

\begin{figure*}[!t]
    \centering
    \subcaptionbox{Labeled Img}{\includegraphics[width=0.19\linewidth]{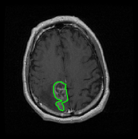}} \hfill
    \subcaptionbox{Level- set}{\includegraphics[width=0.19\linewidth]{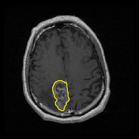}} \hfill
    \subcaptionbox{Our DALS}{\includegraphics[width=0.19\linewidth]{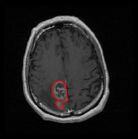}} \hfill
    \subcaptionbox{$\lambda_{1}(x,y)$}{\includegraphics[width=0.19\linewidth]{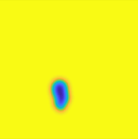}} \hfill
    \subcaptionbox{$\lambda_{2}(x,y)$}{\includegraphics[width=0.19\linewidth]{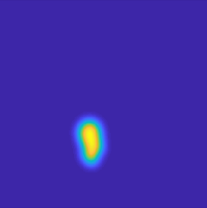}}
    \caption{(a) Labeled image. (b) Level-set (analogous to scalar $\lambda$ parameter constants). (c) DALS output. (d), (e) Learned parameter maps
  $\lambda_{1}(x,y)$ and $\lambda_{2}(x,y)$.}
    \label{fig:lambda_comp}
\end{figure*}

\paragraph{Boundary Delineation:}
As shown in Fig.~\ref{fig:result-comp}, the DALS segmentation contours conform appropriately to the irregular shapes of the lesion boundaries, since the learned parameter maps, $\lambda_1(x,y)$ and $\lambda_2(x,y)$, provide the flexibility needed to accommodate the irregularities. In most cases, the DALS has also successfully avoided local minima and converged onto the true lesion boundaries, thus enhancing segmentation accuracy. DALS performs well for different image characteristics, including low contrast lesions, heterogeneous lesions, and noise. 

\paragraph{Parameter functions and backbone CNN:}
The contribution of the parameter functions was validated by comparing the DALS against a manually initialized level-set ACM with scalar parameters constants as well as with DALS's backbone CNN on its own. As shown in Fig.~\ref{fig:lambda_comp}, the encoder-decoder has predicted the $\lambda_{1}(x,y)$ and $\lambda_{2}(x,y)$ feature maps to guide the contour evolution. The learned maps serve as an attention mechanism that provides additional degrees of freedom for the contour to adjust itself precisely to regions of interest. The segmentation outputs of our DALS and the manual level-set ACM in Fig.~\ref{fig:lambda_comp} demonstrate the benefits of using parameter functions to accommodate significant boundary complexities. Moreover, our DALS outperformed the manually-initialized ACM and its backbone CNN in all metrics across all evaluations on every organ.

\section{Conclusion}

We have presented Deep Active Lesion Segmentation (DALS), a novel framework that combines the capabilities of the CNN and the level-set ACM to yield a robust, fully automatic medical image segmentation method that produces more accurate and detailed boundaries compared to competing state-of-the-art methods. The DALS framework includes an encoder-decoder that feeds a level-set ACM with per-pixel parameter functions. We evaluated our framework in the challenging task of lesion segmentation with a new dataset, MLS, which includes a variety of images of lesions of various sizes and textures in different organs acquired through multiple imaging modalities. Our results affirm the effectiveness our DALS framework.

\bibliography{mlmi19}

\begin{thebibliography}{9}
\providecommand{\natexlab}[1]{#1}
\providecommand{\url}[1]{\texttt{#1}}
\providecommand{\urlprefix}{}

\bibitem[{Chan and Vese(2001)}]{chan2001active}
Chan, T.F., Vese, L.A.: Active contours without edges.
\newblock IEEE Transactions on Image Processing 10(2), 266--277 (2001)

\bibitem[{Hatamizadeh et~al.(2019)Hatamizadeh, Hosseini, Liu, Schwartz, and
  Terzopoulos}]{hatamizadeh2019deep}
Hatamizadeh, A., Hosseini, H., Liu, Z., Schwartz, S.D., Terzopoulos, D.: Deep
  dilated convolutional nets for the automatic segmentation of retinal vessels.
\newblock arXiv preprint arXiv:1905.12120  (2019)

\bibitem[{Hoogi et~al.(2017)Hoogi, Subramaniam, Veerapaneni, and
  Rubin}]{hoogi2017adaptive}
Hoogi, A., Subramaniam, A., Veerapaneni, R., Rubin, D.L.: Adaptive estimation
  of active contour parameters using convolutional neural networks and texture
  analysis.
\newblock IEEE Transactions on Medical Imaging 36(3), 781--791 (2017)

\bibitem[{Hu et~al.(2017)Hu, Shuai, Liu, and Wang}]{hu2017deep}
Hu, P., Shuai, B., Liu, J., Wang, G.: Deep level sets for salient object
  detection.
\newblock In: Proc. IEEE Conf. on Computer Vision and Pattern Recognition
  (2017)

\bibitem[{Imran et~al.(2018)Imran, Hatamizadeh, Ananth, Ding, Terzopoulos, and
  Tajbakhsh}]{hatamizadeh2018automatic}
Imran, A.A.Z., Hatamizadeh, A., Ananth, S.P., Ding, X., Terzopoulos, D.,
  Tajbakhsh, N.: Automatic segmentation of pulmonary lobes using a progressive
  dense {V}-network.
\newblock In: Deep Learning in Medical Image Analysis, Lecture Notes in
  Computer Science, vol. 11045, pp. 282--290. Springer (2018)

\bibitem[{Kass et~al.(1988)Kass, Witkin, and Terzopoulos}]{kass1988snakes}
Kass, M., Witkin, A., Terzopoulos, D.: Snakes: Active contour models.
\newblock International Journal of Computer Vision 1(4), 321--331 (1988)

\bibitem[{Lankton and Tannenbaum(2008)}]{lankton2008localizing}
Lankton, S., Tannenbaum, A.: Localizing region-based active contours.
\newblock IEEE Transactions on Image Processing 17(11), 2029--2039 (2008)

\bibitem[{Marcos et~al.(2018)Marcos, Tuia, Kellenberger, Zhang, Bai, Liao, and
  Urtasun}]{marcos2018learning}
Marcos, D., Tuia, D., Kellenberger, B., Zhang, L., Bai, M., Liao, R., Urtasun,
  R.: Learning deep structured active contours end-to-end.
\newblock In: Proc. IEEE Conf. on Computer Vision and Pattern Recognition
  (CVPR). pp. 8877--8885 (2018)

\bibitem[{Ronneberger et~al.(2015)Ronneberger, Fischer, and
  Brox}]{ronneberger2015u}
Ronneberger, O., Fischer, P., Brox, T.: U-net: Convolutional networks for
  biomedical image segmentation.
\newblock In: LNCS vol. 9351 (Proc. MICCAI). pp. 234--241 (2015)

\end{thebibliography}

\end{document}